\documentclass[a4paper]{article}
\usepackage{graphicx} % Required for inserting images
\usepackage{tabularx}
\usepackage{rotating}
\usepackage{booktabs}
\usepackage{url}
\newcommand{\ra}[1]{\renewcommand{\arraystretch}{#1}}
\newcommand{\na}{\rowcolor{black!5}}
\usepackage{float}
\usepackage{calc}
\usepackage{pdflscape}
\usepackage{colortbl}
\usepackage{longtable}
\usepackage{xcolor}
\usepackage{ragged2e}
\usepackage{pgfplots}
\usepackage{multicol}
\usepackage{stackengine}
\pgfplotsset{width=12cm,compat=1.15}

\usepackage{hyperref}

\title{Survey of Methods, Resources, and Formats for Teaching Constraint Programming}
\author{Tejas Santanam\\Georgia Tech\\Atlanta, Georgia, USA\\\href{mailto:tsantanam@gatech.edu}{tsantanam@gatech.edu} \and Helmut Simonis\footnote{This publication has emanated from research conducted with the financial support of Science Foundation Ireland under Grant number  12/RC/2289-P2 at Insight the SFI Research Centre for Data Analytics at UCC, which is co-funded under the European Regional Development Fund. For the purpose of Open Access, the author has applied a CC BY public copyright licence to any Author Accepted Manuscript version arising from this submission.}\\
Insight SFI Centre for Data Analytics\\School for Computer Science and Information Technology\\University College Cork, Cork, Ireland\\\href{mailto:helmut.simonis@insight-centre.org}{helmut.simonis@insight-centre.org}}
\date{\today}

\begin{document}

\maketitle
\begin{abstract}
    This paper provides an overview of the state of teaching for Constraint Programming, based on a survey of the community for the 2023 Workshop on Teaching Constraint Programming at the CP 2023 conference in Toronto. The paper presents the results of the survey, as well as lists of books, video courses and other tutorial materials for teaching Constraint Programming. The paper serves as a single location for current and public information on course resources, topics, formats, and methods.
\end{abstract}

\section{Introduction}
\label{sec:introduction}

Constraint Programming (CP) is a declarative problem solving paradigm that has been used over the last forty years to solve many industrial combinatorial problems, and that can be used to express and solve problems with many different problem solving technologies and methods. The approach to express and solve a problem with CP is quite different to the programming in a traditional programming language that most computer science students are familiar with, therefore training of students in CP techniques is quite important for the area. 

After a successful workshop on Teaching Constraint Programming at the CP conference in Cork in 2015, organized by Patrick Prosser, another workshop on this topic was organized for CP 2023 in Toronto \footnote{https://hsimonis.github.io/WTCP2023/}. The papers and presentations from that workshop can be found \href{https://hsimonis.github.io/WTCP2023/}{here}. Many new CP systems and courses have appeared since 2015, and the Covid-19 pandemic has led to many, perhaps only temporary, changes in the way students are taught. As part of the workshop program, it was decided to create a survey of the community on this topic. This paper describes the survey, its results, and possible future actions intended to improve teaching for CP. This paper is therefore intended to first present information about courses being run, and materials being used, then to identify gaps in the currently available materials, and then to suggest possible types of courses for development. 

While we received 51 replies to the survey, and had further feedback on the topic during the discussions at the workshop, we are probably still missing important information about courses held at other institutions, or in other settings. The paper should be therefore seen as a working document, with new links and materials being added over time.

The paper is structured as follows: We first (Section~\ref{sec:survey}) describe the survey we ran as part of the Workshop on Teaching Constraint Programming in 2023, describing the methodology in Section~\ref{sec:Methodoloy}, then presenting the results in Section~\ref{sec:results}, and listing additional comments made by the participants in Section~\ref{sec:comments}, before drawing some conclusions of our analysis of the results in Section~\ref{sec:analysis}. In the next Section~\ref{sec:coursedescriptions} we list the course description links we have received as part of the survey results. Most of the links point to a description of the course only, but some include slides sets or videos. We discuss course materials for such freely available courses in more detail in Section~\ref{sec:coursematerials}. The next Section, Section~\ref{sec:books}, lists books on Constraint Programming which are used for teaching CP. This is followed by a description of video courses in Section~\ref{sec:videocourses}, and of the existing literature on teaching of CP in Section~\ref{sec:articles}.  We have also included a section on tutorials (Section~\ref{sec:tutorials}) and on additional content (Section~\ref{sec:othermaterial}) that can be helpful in developing teaching materials. We discuss potential course structures for further development in Section~\ref{sec:cpcourseformats}, before concluding the paper in Section~\ref{sec:conclusions}.

\section{Survey}
\label{sec:survey}

As part of organizing the Workshop on Teaching Constraint Programming for CP 2023 we decided to try and poll the community about where and how CP is taught. For this purpose, we designed an \href{https://forms.gle/v54HUsbSXcyHmfME9}{on-line questionnaire} with Google Forms, and then invited the community to participate. 

There were three main points we wanted to study. 
\begin{itemize}
    \item First, who is teaching Constraint Programming, is this purely happening at universities, or are there other ways of learning about CP?
    \item Can we determine common topics and areas covered, or are the courses widely different from each other? Is there a consensus about the tools to be used?
    \item Can we collect teaching resources that might be shared/reused by other institutions?
\end{itemize} 

In particular, we sought answers to the following questions: 

\begin{itemize}
    \item Where are CP courses taught, who is the target audience, how many students are reached? Are there regional gaps where CP is not visible?
    \item If a CP researcher is not teaching a course on CP, what are the reasons?
    \item Are these courses well established, or are they a rather recent introduction to the teaching program?
    \item How long is the course?
    \item Is there a textbook that is widely used? If not, what other resources are used?
    \item Which topics are included in the course, which areas are left out?
    \item Do the students get hands-on experience with a CP system, if yes, which system is used?
    \item What type of problems do the students solve?
    \item After the experience during the Covid-19 pandemic, have universities returned to a traditional teaching format?
    \item Has there been an impact of the growing popularity of large language models (LLM)?
\end{itemize}

\subsection{Methodology}
\label{sec:Methodoloy}

% \emph{explain why we did ask these questions, and not others}

Table~\ref{tab:questions} lists the questions asked in the survey, the format of the answers, and a link to the presentation of the answers.

{\scriptsize
\begin{longtable}{rp{7cm}rr}
\caption{\label{tab:questions}Survey Questions}\\ 
\toprule
Nr & Question & \shortstack[l]{Reply\\Format} & Answers\\ \midrule
\endfirsthead
\caption{Survey Questions (cont'd)}\\ 
\toprule
Nr & Question & \shortstack[l]{Reply\\Format} & Answers\\ \midrule
\endhead
\bottomrule
\endfoot
1 & Does your institution offer a CP course or a course that covers some content around CP, SAT, or similar? & yes/no & Figure \ref{fig:offer}\\
2 & If no, what are the barriers or reasons? & text& Page~\pageref{barriers}\\
3 & If your institution offers such a course, is there more than one course that covers some aspect of this topic? (added during the survey)& yes/no & Answers incomplete\\
4 & If you answered yes to the first question, who is the audience? & text& Figure \ref{fig:audience}\\
5 & How long has the course been taught? & Categories & Figure \ref{fig:howlong}\\
6 & How many students does each offering of the class have? & Categories & Figure \ref{fig:howmanystudents}\\
7 & How many hours of instruction does the course involve in total? & Categories& Figure \ref{fig:hours} \\
8 & Does the class use a textbook or similar resource? & yes/no & Figure \ref{fig:textbook}\\
9 & If you answered yes to the above question, list the resource below & text& Figure \ref{fig:resources} \\
10 & Does the class involve coding? & yes/no & Figure \ref{fig:coding}\\
11 & If you answered yes to the above question, list the language/software below & text& Figures \ref{fig:solver}, \ref{fig:language} \\
12 & Have you seen any impact from Large Language Models (LLMs) such as ChatGPT in the course? & text& Figure \ref{fig:llm}\\
13 & Is the instruction primarily delivered through in-person interaction or through videos/MOOC? & text& Figure \ref{fig:mooc}\\
14 & What sort of exercises do the students solve within the context of the course? & text & Figures \ref{fig:exerciseareas},\ref{fig:exercisetypes}\\
15 & Are there any public links you can provide on the course (ranging from course description to course materials)? & text& Section \ref{sec:coursedescriptions}\\
16 & Any other comments that you would like to make? & text& Section~\ref{sec:comments}\\
\end{longtable}
}

\subsection{Results}
\label{sec:results}

We now present the results of the survey, describing the answers to each question in turn.

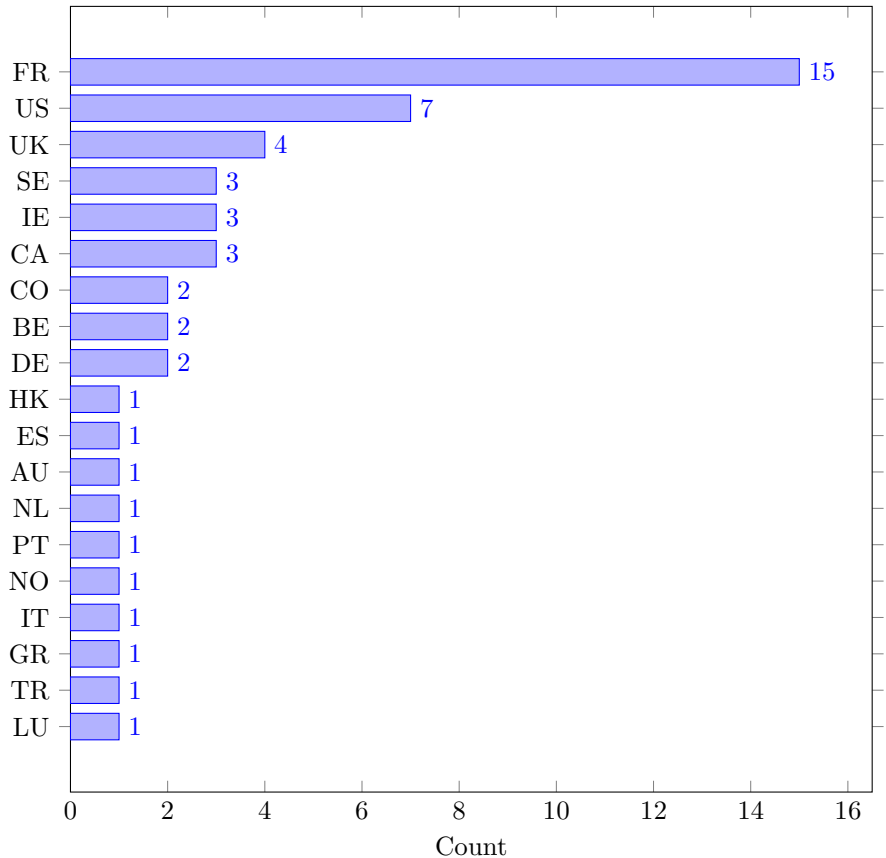
\begin{figure}[htbp]
\centering
\begin{tikzpicture}
\begin{axis}[xbar,symbolic y coords={LU,TR,GR,IT,NO,PT,NL,AU,ES,HK,DE,BE,CO,CA,IE,SE,UK,US,FR},width=\textwidth,height=12cm,xmin=0,ytick=data,xlabel=Count,nodes near coords, nodes near coords align={horizontal}]
\addplot coordinates{(15,FR)(7,US)(4,UK)(3,SE)(3,IE)(3,CA)(2,CO)(2,BE)(2,DE)(1,HK)(1,ES)(1,AU)(1,NL)(1,PT)(1,NO)(1,IT)(1,GR)(1,TR)(1,LU)};
\end{axis}
\end{tikzpicture}
\caption{\label{fig:bycountry}Participants By Country}
\end{figure}

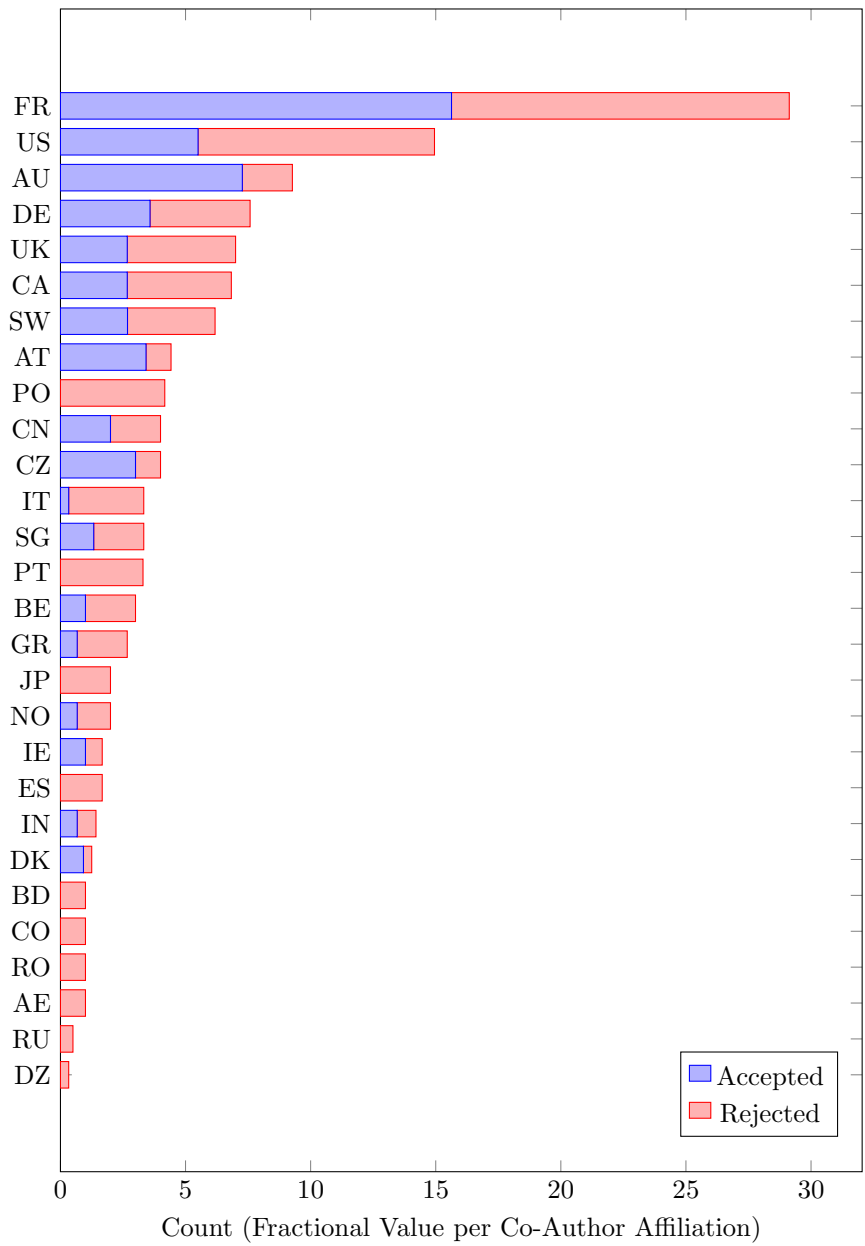
\begin{figure}[htbp]
\begin{tikzpicture}
\begin{axis}[xbar stacked,
symbolic y coords={DZ,RU,AE,RO,CO,BD,DK,IN,ES,IE,NO,JP,GR,BE,PT,SG,IT,CZ,CN,PO,AT,SW,CA,UK,DE,AU,US,FR},
width=\textwidth,
height=17cm,
xmin=0,
ytick=data,
xlabel=Count (Fractional Value per Co-Author Affiliation),
legend pos=south east
%nodes near coords, nodes near coords align={horizontal}
]
 \addplot coordinates{(15.63,FR)(5.5,US)(7.27,AU)(3.58,DE)(2.67,UK)(2.67,CA)(2.68,SW)(3.42,AT)(0,PO)(2,CN)(3,CZ)(0.33,IT)(1.33,SG)(0,PT)(1,BE)(0.67,GR)(0,JP)(0.67,NO)(1,IE)(0,ES)(0.67,IN)(0.92,DK)(0,BD)(0,CO)(0,RO)(0,AE)(0,RU)(0,DZ)}; 
 
\addplot coordinates{(13.5,FR)(9.45,US)(2,AU)(4,DE)(4.33,UK)(4.16,CA)(3.5,SW)(1,AT)(4.17,PO)(2,CN)(1,CZ)(3,IT)(2,SG)(3.3,PT)(2,BE)(2,GR)(2,JP)(1.33,NO)(0.67,IE)(1.67,ES)(0.75,IN)(0.33,DK)(1,BD)(1,CO)(1,RO)(1,AE)(0.5,RU)(0.33,DZ)};  
\legend{\strut Accepted,\strut Rejected};
\end{axis}
\end{tikzpicture}

\caption{\label{fig:CP2020}Accepted/Rejected Papers by Country for the CP 2020 conference~\cite{DBLP:conf/cp/2020}}
\end{figure}

Figure~\ref{fig:bycountry} shows the participants per country, the largest number of participants comes from France, followed by the US, and the UK. Many other countries have three or fewer participants. Notable is the lack of entries from Asia (excepting Hong Kong), Africa, and South America (excepting Colombia with two entries from the same university). While the overall numbers seem roughly in line with submission numbers at the CP conference (see Figure~\ref{fig:CP2020} for the submitted/accepted papers for the CP2020 conference~\cite{DBLP:conf/cp/2020}), it points to a lack of CP education in large parts of the globe. We need to investigate further if these gaps are due to lack of awareness of the survey, or, more seriously, a lack of CP interest in these areas.

Figure~\ref{fig:institutions} shows the participants by country and institution. Some institutions had multiple participants, entering different courses, so the total number of entries does not match the total number of participants. Note that some well-known CP centres are missing (e.g. Vienna, Singapore, Barcelona), we should look for feedback from these centres.  

\begin{figure}[htbp]
{\scriptsize
\begin{multicols}{2}
\begin{itemize}
\item France
\begin{itemize}
\item  CRIL Lens
\item  Montpellier
\item  Grenoble
\item  Sophia Antipolis
\item  Nice
\item  Cote Azur
\item  Ecole Polytechnique
\item  IMT Atlantique
\item  INRAE
\item  UPHF Valenciennes 
\item  Cosling
\item  IRISA, Rennes
\item  INRIA
\end{itemize}

\item USA
\begin{itemize}
\item  UT Dallas
\item  Brown
\item  CMU
\item  UConn
\item  City University New York
\item  Georgia Tech
\item  Lincoln, Nebraska
\end{itemize}

\item Canada
\begin{itemize}
\item  Laval
\item  Poly Montreal
\item  Toronto
\end{itemize}

\item UK
\begin{itemize}
\item  Edinburgh
\item  York
\item  St Andrews
\item  Glasgow
\end{itemize}

\item Sweden
\begin{itemize}
\item  Lund Univ
\item  RISE
\item  Upssala Univ
\end{itemize}

\item Belgium
\begin{itemize}
\item  UCLouvain
\item  KU Leuven
\end{itemize}
\item Germany
\begin{itemize}
\item  Fraunhofer
\item  TU Cottbus
\end{itemize}

\item Ireland
\begin{itemize}
\item  Munster Technical University
\item  University College Cork
\end{itemize}

\item Others
\begin{itemize}
\item  UDG Girona, Spain
\item  CUHK, China
\item  Monash, Australia
\item  Delft, The Netherlands
\item  Lisbon, Portugal
\item  Bologna, Italy
\item  Simula, Norway
\item  Western Macedonia, Greece
\item  Universidad del Valle, Colombia
\item  Izmir,Turkey
\item  Univ. Luxembourg, Luxembourg
\end{itemize}
\end{itemize}
\end{multicols}
}
\caption{\label{fig:institutions}Participants by Institution}
\end{figure}

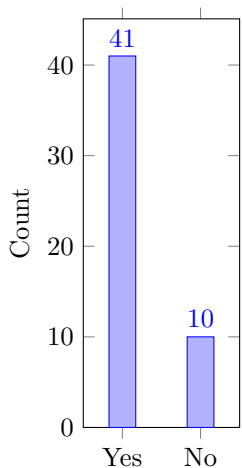
\begin{figure}[htbp]
\centering
\begin{tikzpicture}
\begin{axis}[ybar,symbolic x coords={Yes,No},width=.3\textwidth,height=7cm,ymin=0,xtick=data,ylabel=Count,nodes near coords, nodes near coords align={vertical},enlarge x limits=0.5]
\addplot coordinates{(Yes,41)(No,10)};
\end{axis}
\end{tikzpicture}
\caption{\label{fig:offer}Does your Institution offer a CP Course?}
\end{figure}

Not all survey participants run a CP course at their institutions, 10 out of 51 participants answered "No" in Figure~\ref{fig:offer}.

The reasons given vary, some of the reasons stated are: \label{barriers}

\begin{itemize}
\item (4) No students (industry, research centre)
\item (2) Time/Workload
\item (2) Only as part of a more general course
\item (1) No interest by students
\item (1) Not allowed by institution
\item (1) CP is not a recommended course in the ACM Curricula Recommendations. 
\end{itemize} 
We did receive one reply describing courses run by industry. These courses are run on-demand for professionals, using proprietary teaching materials on modelling industrial problems.

\begin{figure}[htbp]
\centering
\begin{tikzpicture}
\begin{axis}[ybar,symbolic x coords={Undergrad, Graduate,Masters,PhD students},width=\textwidth,height=7cm,ymin=0,xtick=data,ylabel=Count,nodes near coords, nodes near coords align={vertical},enlarge x limits=0.5]
\addplot coordinates{(Undergrad, 20)(Graduate,34)(Masters,3)(PhD students,1)};
\end{axis}
\end{tikzpicture}
\caption{\label{fig:audience}Intended Audience of Course}
\end{figure}
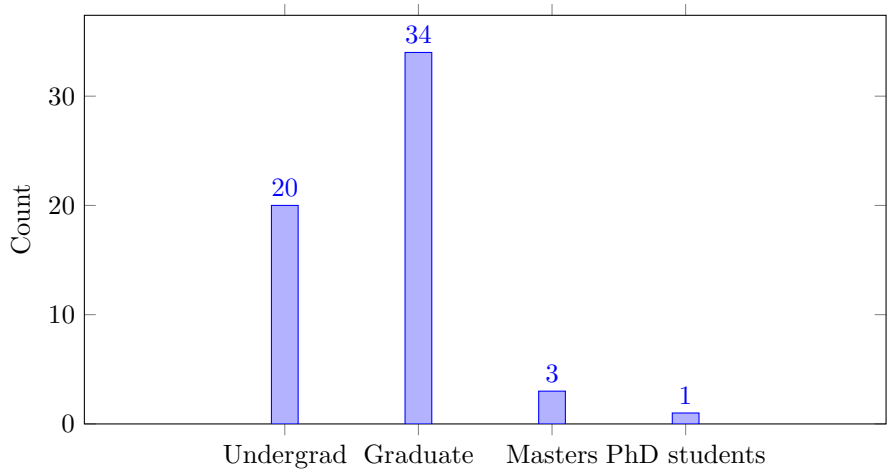

Figure~\ref{fig:audience} shows the intended audience for the courses. Some institutions offer courses at the undergraduate and graduate level, while others allow both undergraduates and graduate students to attend the same course. The total number of entries therefore does not match the number of participants running courses. One course (\cite{Choi2023}) is provided for PhD students as part of a general, national training program for AI in Ireland.

\begin{figure}[htbp]
\centering
\begin{tikzpicture}
\begin{axis}[ybar,symbolic x coords={10+ years,5-10 years,3-5 years,1-3 years,0-1 years,starting},width=\textwidth,height=7cm,ymin=0,xtick=data,ylabel=Count,nodes near coords, nodes near coords align={vertical}]
\addplot coordinates{(10+ years,20)(5-10 years,9)(3-5 years,12)(1-3 years,4)(0-1 years,0)(starting,1)};
\end{axis}
\end{tikzpicture}
\caption{\label{fig:howlong}How long has the course been offered?}
\end{figure}
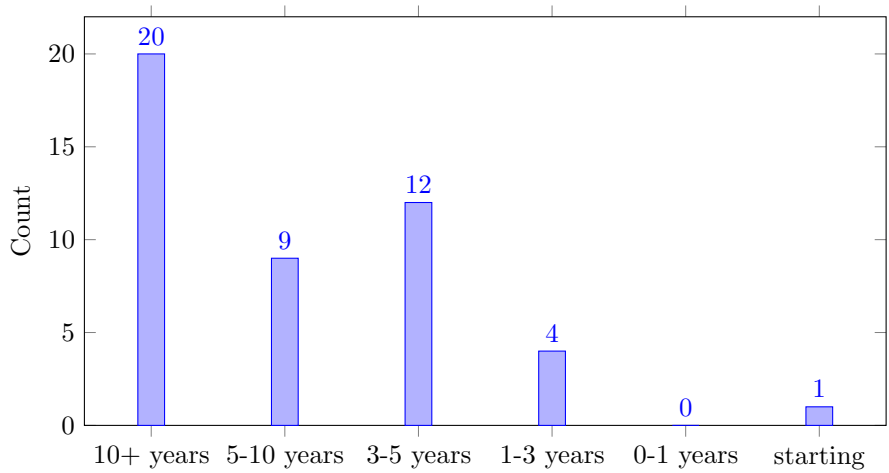

Many of the courses have been running for many years, as Figure~\ref{fig:howlong} shows. Some participants commented that while a course has been running for multiple years, the content of the course has changed over time, adapting different tools and objectives. On the other hand, four courses have been running for three or fewer years, and one course is starting this year.

Enrollment numbers vary quite significantly. Some courses are run with ten or fewer students, while two are marked as having 100+ students.
%\footnote{We still have to verify that this is the number of participants per run of the course, not the total over all years. (Lund)}.

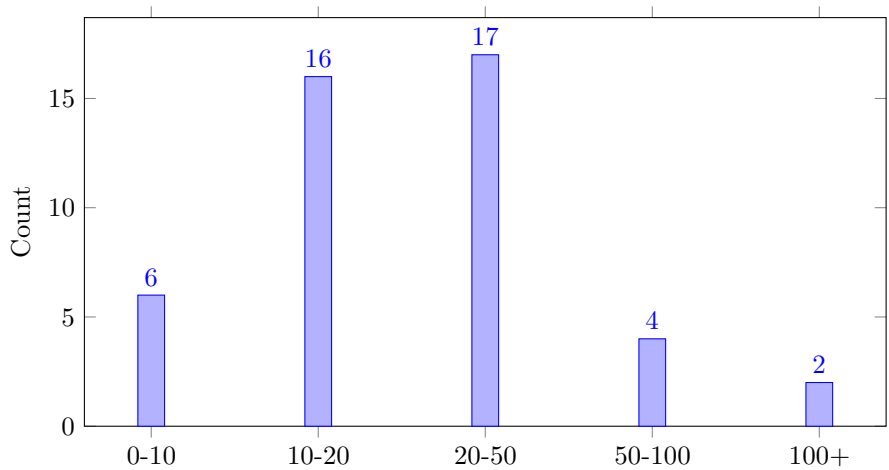
\begin{figure}[htbp]
\centering
\begin{tikzpicture}
\begin{axis}[ybar,symbolic x coords={0-10,10-20,20-50,50-100,100+},width=\textwidth,height=7cm,ymin=0,xtick=data,ylabel=Count,nodes near coords, nodes near coords align={vertical}]
\addplot coordinates{(0-10,6)(10-20,16)(20-50,17)(50-100,4)(100+,2)};
\end{axis}
\end{tikzpicture}
\caption{\label{fig:howmanystudents}How many students does each offering of the course have?}
\end{figure}

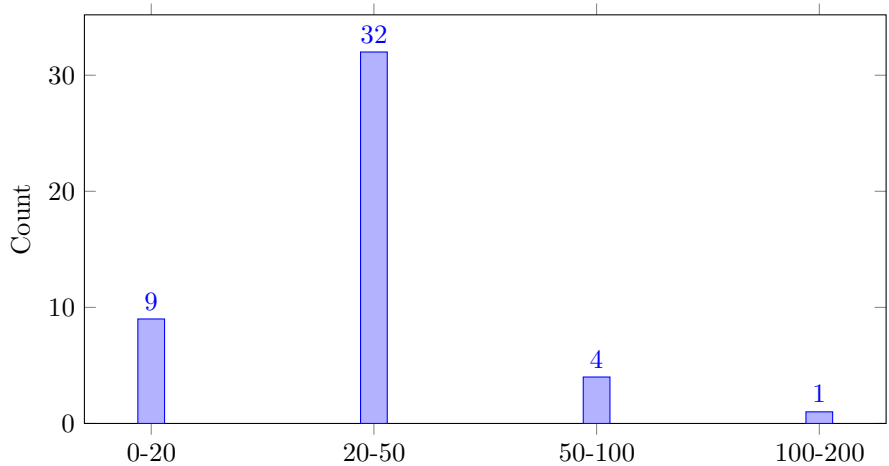
\begin{figure}[htbp]
\centering
\begin{tikzpicture}
\begin{axis}[ybar,symbolic x coords={0-20,20-50,50-100,100-200},width=\textwidth,height=7cm,ymin=0,xtick=data,ylabel=Count,nodes near coords, nodes near coords align={vertical}]
\addplot coordinates{(0-20,9)(20-50,32)(50-100,4)(100-200,1)};
\end{axis}
\end{tikzpicture}
\caption{\label{fig:hours}How many hours of instruction does the course have?}
\end{figure}

The majority of the courses offered provide between 20 and 50 hours of instruction, as Figure~\ref{fig:hours} shows. One course provides more than 100 hours of instruction.

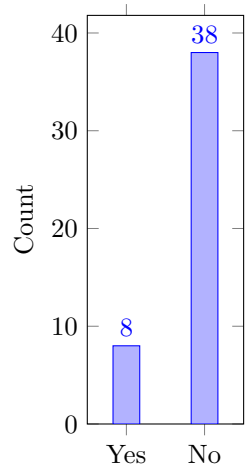
\begin{figure}[htbp]
\centering
\begin{tikzpicture}
\begin{axis}[ybar,symbolic x coords={Yes,No},width=.3\textwidth,height=7cm,ymin=0,xtick=data,ylabel=Count,nodes near coords, nodes near coords align={vertical},enlarge x limits=0.5]
\addplot coordinates{(Yes,8)(No,38)};
\end{axis}
\end{tikzpicture}
\caption{\label{fig:textbook}Does the class use a textbook or similar resource?}
\end{figure}

Figure~\ref{fig:textbook} shows that relatively few (8) courses use a textbook, while most do not. 

The participants were asked which resources were used, the answer shows that there is little commonality in resources used. Each course largely depends on its own materials. We will discuss textbooks in more detail in Section~\ref{sec:books}. Figure~\ref{fig:resources} lists the resources mentioned.
\begin{figure}[htbp]
{\scriptsize
\begin{itemize}
\item (2) Krzysztof Apt "Principles of Constraint Programming" \cite{DBLP:books/daglib/0018273}
\item P. Hofstedt, A. Wolf "Einf{\"{u}}hrung in die Constraint-Programmierung"\cite{DBLP:books/daglib/0034544}
\item R. Dechter "Constraint processing" \cite{DBLP:books/daglib/0016622}
\item Jupyter notebooks
\item MiniZinc Handbook
\item EdX MOOC
\item MiniCP material
\item Krzysztof Kuchcinski, "Modeling and Optimization of Embedded System with Constraint Programming: Principles and Practice" \footnote{I did not find a description of this book?}
\item Global Constraint Catalog~\cite{DBLP:journals/constraints/BeldiceanuCDP07}
\item Choco-solver manual
\item Research papers
\end{itemize}
}
\caption{\label{fig:resources}Resources}
\end{figure}

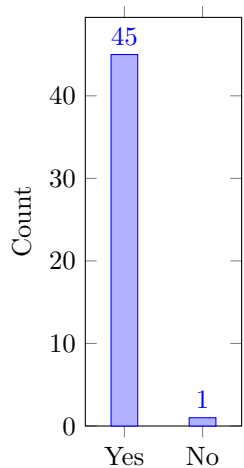
\begin{figure}[htbp]
\centering
\begin{tikzpicture}
\begin{axis}[ybar,symbolic x coords={Yes,No},width=.3\textwidth,height=7cm,ymin=0,xtick=data,ylabel=Count,nodes near coords, nodes near coords align={vertical},enlarge x limits=0.5]
\addplot coordinates{(Yes,45)(No,1)};
\end{axis}
\end{tikzpicture}
\caption{\label{fig:coding}Does the class involve coding?}
\end{figure}

Nearly all courses involve coding (Figure~\ref{fig:coding}), only one respondent answered "No".%\footnote{To be verified, E. Bourreau}.

\begin{figure}[htbp]
\centering
\begin{tikzpicture}
\begin{axis}[ybar,symbolic x coords={MiniZinc,Choco,MiniCP,CP Opt,Any,Others},width=\textwidth,height=7cm,ymin=0,xtick=data,ylabel=Count,nodes near coords, nodes near coords align={vertical}]
\addplot coordinates{(MiniZinc,20)(Choco,9)(MiniCP,5)(CP Opt,3)(Any,5)(Others,20)};
\end{axis}
\end{tikzpicture}
\caption{\label{fig:solver}Solver/System Used? (Multiple Allowed)}
\end{figure}
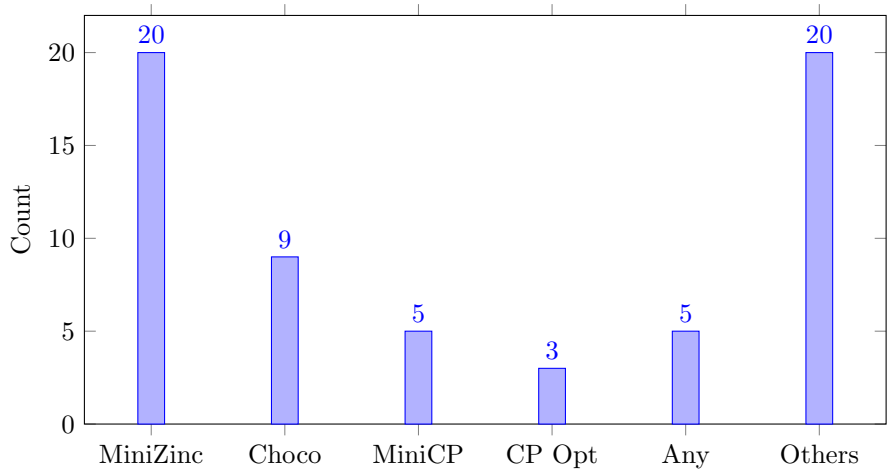

\begin{figure}[htbp]
\centering
\begin{tikzpicture}
\begin{axis}[ybar,symbolic x coords={Java,Python,Prolog,C++,Any,Others},width=\textwidth,height=7cm,ymin=0,xtick=data,ylabel=Count,nodes near coords, nodes near coords align={vertical}]
\addplot coordinates{(Java,11)(Python,8)(Prolog,5)(C++,4)(Any,5)(Others,2)};
\end{axis}
\end{tikzpicture}
\caption{\label{fig:language}Programming Language Used? (Multiple Allowed)}
\end{figure}
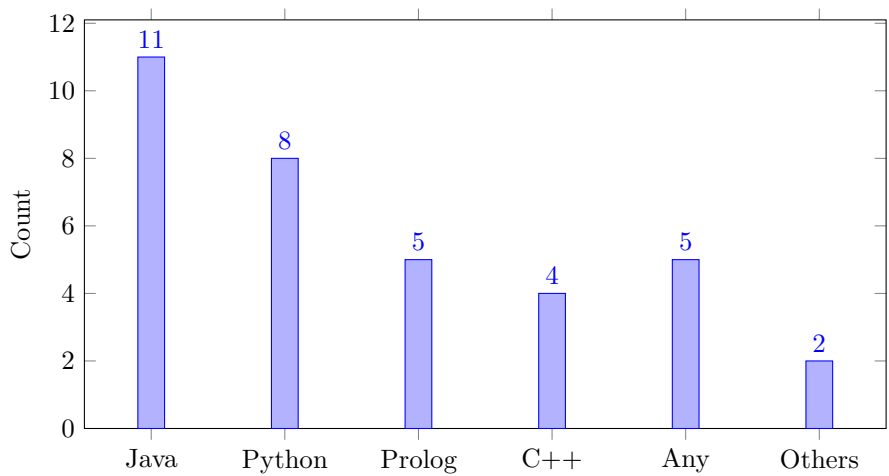

We asked which CP system and programming language were used in the course, multiple answers were possible in a free-form field. Figure~\ref{fig:solver} shows the answers for the CP system, Figure~\ref{fig:language} shows the answers for the programming language used. In Figure~\ref{fig:solver}, the largest number of replies mentioned MiniZinc~\cite{DBLP:conf/cp/NethercoteSBBDT07} (the back-end solver used was not mentioned in most entries, but Gecode~\cite{SchulteTackLagerkvist:MPG:2010} was mentioned several times). This was followed in frequency by Choco-solver~\cite{DBLP:journals/jossw/PrudhommeF22} and MiniCP~\cite{MiniCP2021}, which is rather interesting given that its release is quite recent. Five replies stated that the students can work with any CP system they like, while a large number of system were only listed in one reply.

The following systems were mentioned by only one participant:
\begin{multicols}{3}
\begin{itemize}
\item SICStus Prolog~\cite{carlsson2010sicstus}
\item SAT
\item Conjure~\cite{DBLP:conf/ijcai/AkgunFGJMN23}
\item Savile Row~\cite{DBLP:journals/corr/abs-2201-03472}
\item PyCSP3~\cite{DBLP:journals/corr/abs-2009-00326}
\item ACE~\cite{DBLP:journals/corr/abs-2302-05405}
\item pysat
\item Z3
\item Clingo
\item OPL~\cite{DBLP:journals/informs/Hentenryck02}
\item Cplex
\item Hava
\item JaCoP~\cite{JaCoP2013}
\item SWI Prolog~\cite{wielemaker:2011:tplp}
\item Essence'
\item Torch
\item pytoulbar2
\item ECLiPSe~\cite{DBLP:journals/tplp/SchimpfS12}
\item OR-Tools
\item Gecode~\cite{SchulteTackLagerkvist:MPG:2010}
\end{itemize}
\end{multicols}

The most common programming language used (Figure~\ref{fig:language}) is Java, followed by Python, and different versions of Prolog (ECLiPSe~\cite{DBLP:journals/tplp/SchimpfS12}, SICStus Prolog~\cite{carlsson2010sicstus} and SWI Prolog~\cite{wielemaker:2011:tplp}), and finally C++. In five replies it is stated that the students can pick any language they like. Note that MiniZinc is normally used with its own IDE, without using a host language, but that some answers mentioned the use of Python to host the MiniZinc back-end solver.
The following languages were mentioned once each.
\begin{itemize}
\item Julia
\item Scala
\end{itemize}

\begin{figure}[htbp]
\centering
\begin{tikzpicture}
\begin{axis}[ybar,symbolic x coords={In-person, Virtually, Mixed, Flipped},width=\textwidth,height=7cm,ymin=0,xtick=data,ylabel=Count,nodes near coords, nodes near coords align={vertical}]
\addplot coordinates{(In-person,41)(Virtually,2)(Mixed,1)(Flipped,2)};
\end{axis}
\end{tikzpicture}
\caption{\label{fig:mooc}Is the instruction primarily delivered through in-person
interaction or through videos/MOOC?}
\end{figure}
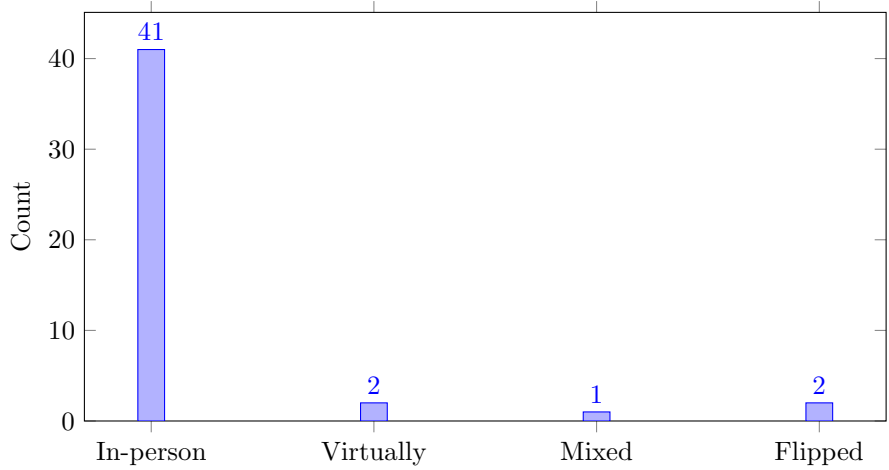

The last question was about the delivery of the courses. During the Covid-19 pandemic, many courses had to use a virtual format, but this has not been continued after the restrictions were lifted. 
The vast majority of the courses is still given in-person, as shown in Figure~\ref{fig:mooc}. Some courses use a flipped format, where lectures are pre-recorded, and are available on-line, while the direct interaction with the students is used to answer questions, and provide feedback.

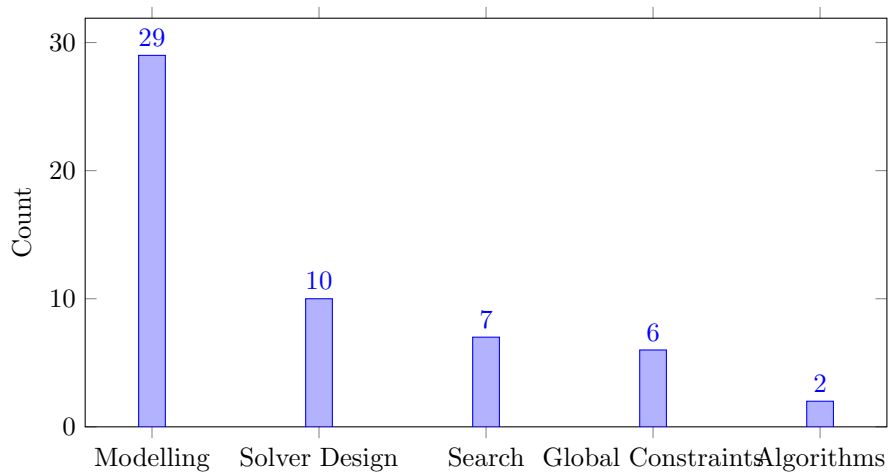
\begin{figure}[htbp]
\centering
\begin{tikzpicture}
\begin{axis}[ybar,symbolic x coords={Modelling,Solver Design,Search,Global Constraints,Algorithms},width=\textwidth,height=7cm,ymin=0,xtick=data,ylabel=Count,nodes near coords, nodes near coords align={vertical}]
\addplot coordinates{(Modelling,29)(Solver Design,10)(Search,7)(Global Constraints,6)(Algorithms,2)};
\end{axis}
\end{tikzpicture}
\caption{\label{fig:exerciseareas}What sort of exercises do the students solve (Free Form, Areas)?}
\end{figure}

\begin{figure}[htbp]
\centering
\begin{tikzpicture}
\begin{axis}[ybar,symbolic x coords={Puzzles,Scheduling,Logistics,Time Tabling,Planning},width=\textwidth,height=7cm,ymin=0,xtick=data,ylabel=Count,nodes near coords, nodes near coords align={vertical}]
\addplot coordinates{(Puzzles,7)(Scheduling,7)(Logistics,4)(Time Tabling,2)(Planning,1)};
\end{axis}
\end{tikzpicture}
\caption{\label{fig:exercisetypes}What sort of exercises do the students solve (Free Form, Types)?}
\end{figure}
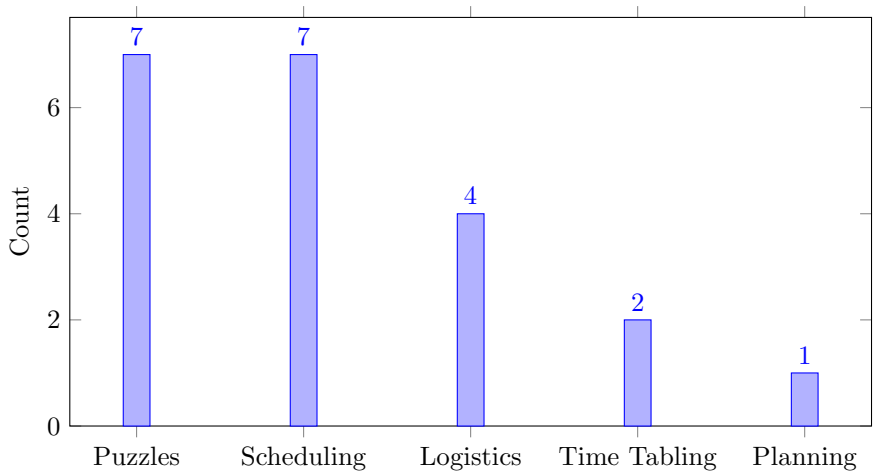

\subsection{Additional Comments}
\label{sec:comments}

We provided a free form field to provide additional feedback. Several of the participants noted that they teach CP as part of a larger course, we probably did not consider the implications of that in the design of the questionnaire carefully enough.

Another comment was that in multiple places there are more than one course on CP, again the design of the survey did not allow for that.

More specific comments were:
\begin{quote}
    The course is pretty challenging for students without a lot of computational thinking, and since it doesn't have many direct prerequisites there are always a number of weak students who struggle a lot. Partly because we have a generation of weak students who complete assignments by a combination of LLM and reddit and trial and error!
\end{quote}

\begin{quote}
    Students who take the course love it.
\end{quote}

\begin{quote}
    When they start, they have had a Prolog class and they think Prolog is difficult. At the end of the CP class, they think Prolog is quite easy compared to constraints. We would need a 3rd LP course... 
\end{quote}

The student experience seems to differ widely depending on the topics included in the course, the background of the students, and the overall teaching aims. But a common comment was that students struggle with the declarative nature of CP, perhaps more so if they are only used to writing imperative programs.

\begin{quote}
    This is a timely workshop because the ACP EC is currently discussing the dissemination of CP through turnkey teaching modules that can be incorporated into university courses and through turnkey tutorials that can be proposed in related conferences.
\end{quote}

There seems to be a need for different course formats, from short overviews that can be easily included in another course, to modelling courses on how to use CP, to more detailed engineering courses on how to write CP solvers.

\begin{quote}
    Prolog with appropriate constraint solving and constraint-based modeling libraries should be the right language to teach CP (unlike MiniZinc for several reasons: explicit logical foundations, constraints as answers not just instantiated solutions, possibility to program constraint solvers without changing of language, no error message following transformation to FlatZinc)
\end{quote}
Clearly, people are passionate about their specific approach to presenting Constraint Programming. This seems to indicate that it might be difficult to come up with a single, uniform course format that satisfies everybody.

\begin{quote}
    Setting up a shared resource of teaching materials would be great.
\end{quote}

\begin{quote}
    And thank you for your initiative collecting this information.  Big service to the community.
\end{quote}

\begin{quote}
    I cannot attend CP but I would like to have a summary of what you discussed.
\end{quote}

On the other hand, there clearly is an interest in shared resources, and providing some information how other people teach CP.

\subsection{Analysis}
\label{sec:analysis}

It is important to note that we are trying to generalize from a relatively small sample. There were 51 responses to the survey, we don't know how many places that do teach CP are not included. Given that caveat, the following conclusions can be drawn:

\begin{enumerate}
    \item There are large areas of the world (especially many parts of Asia, Africa, and South America) where Constraint Programming is not taught. It should be one of the objectives of the ACP to remedy this.
    \item CP is almost exclusively taught at universities, mostly at the advanced undergraduate and graduate level. We did not receive replies from industry about in-house courses, nor third party courses being used.
    \item In universities where CP is not taught, the reasons given are time needed to develop a course, little interest from students, or from the university itself. In many of these cases, providing shared resources through the ACP might be helpful to overcome these difficulties. 
    \item Non-traditional courses, like MOOC (Massive Open Online Courses), have reached a much wider audience than most CP courses. Some exceptional courses at universities reach more than 100 students per instance.
    \item Many existing CP courses have been running for many years, showing that the topic is well established at their university. 
    \item A course format of a one-semester course (approximately 24-40 hours taught) is quite common, but CP is also taught as a sub-module in more general courses, or as a significantly longer course.
    \item There currently is no single, standard textbook that is widely used. Most courses rely on their own resources, and using tutorials and manuals provided by the CP systems used.
    \item Nearly all courses involve student exercises through coding, mostly focused on modelling simple problems. The use of puzzles is widespread, as well as using scheduling and logistics examples.
    \item There is little consensus on which CP system should be used, but MiniZinc is used for many courses that focus on modelling. Many tools mentioned are only used in a single location.
    \item Most universities have reverted to a standard in-person course format after the disruption caused by the Covid-19 pandemic.
    \item At the time of the survey, the impact of large language models (LLM) on the courses was still quite limited, but a significant impact is expected by many participants.
\end{enumerate}

\clearpage
\section{Course Descriptions}
\label{sec:coursedescriptions}

Table~\ref{tab:coursedescriptions} gives a more detailed overview of the courses taught at different universities which were mentioned in the survey responses. The table lists them in order of reply, we list the country and name of the organization providing the course, the language in which is course is taught, and the CP system that is used. If that is not provided in the available material, the entry is left blank. We also list the indented audience, and the format of the information provided. Possible values are \textbf{CD} for a course description, \textbf{WB} for workbook, \textbf{Slides}, if the slides of the course are available at the link given, or \textbf{Video}, if a video course is described. The last column gives the name of the course, together with a link to a web page which provides the information. Some links might get stale quite quickly, as the descriptions often are linked to a specific semester or year when the course is taught. We see that the majority of the courses are given in English, but there may be versions of the material in other languages, e.g. French, German or Italian, if a course is taught in different languages depending on the student group. 

For some organizations, like \href{https://ucc-ie-public.courseleaf.com/modules/}{UCC} or \href{https://docs.google.com/spreadsheets/d/1kZt5TLQ7VFotJbuEDRAXSFoG-Y3U-rFAB8bYenpMP9Y/edit#gid=74570599}{Montpellier}, the link to the course description goes to a higher-level curriculum overview, and you have to find the relevant entries with the name of the course provided.  

\begin{landscape}
\begin{longtable}{lp{4cm}p{1.5cm}p{2cm}llp{9cm}}
\caption{\label{tab:coursedescriptions}Course Descriptions}\\ \toprule
Country & Organization & Language & CP System & Audience & Format & Course Name/Link \\ \midrule
\endfirsthead
\caption{Course Descriptions}\\ \toprule
Country & Organization & Language & CP System & Audience & Format & Course Name/Link \\ \midrule
\endhead
US &UT Dallas& English & Prolog & & CD& \href{https://utdallas.edu/~gupta/courses/lp/}{Computational Logic} \\
UK& St Andrews& English & & & CD &\href{https://www.st-andrews.ac.uk/subjects/modules/catalogue/?code=CS4402&academic_year=2023%2F4}{Constraint Programming} \\
HK &Chinese University, Hong Kong& English & &  & CD& \href{https://www.cse.cuhk.edu.hk/~jlee/courses/flippedCSCI5240.html}{Combinatorial Search and Optimization with Constraints} \\
US &Brown University& English & IBM Optimization Studio& & CD&  \href{https://cs.brown.edu/courses/csci2951-o/index.html}{Foundations of Prescriptive Analytics} \\
FR &University Artois& English & & & CD & \href{https://www.cril.univ-artois.fr/~lecoutre/#/teaching}{Constraint Programming} \\
BE &KU Leuven & English & & & CD& \href{https://onderwijsaanbod.kuleuven.be//2022/syllabi/e/H02A8AE.htm#activetab=doelstellingen_idp163248}{Advanced Programming Languages for A.I.} \\
CA & University Laval& French & & & Slides& \href{http://www2.ift.ulaval.ca/~quimper/Optimisation/Diapositives/chapitres.php}{Optimization} \\
CA & Polytechnique Montreal & French  & & & CD & \href{https://www.polymtl.ca/programmes/cours/programmation-par-contraintes}{Programmation par contraintes} \\
SE& Uppsala University & English & MiniZinc, MiniCP& Masters& CD & \href{https://www.uu.se/en/study/course?query=1DL442}{Combinatorial Optimisation and Constraint Programming} \\
& & English & MiniZinc, Gecode& Masters & Slides& \href{https://user.it.uu.se/~pierref/courses/COCP/slides/}{Combinatorial Optimisation and Constraint Programming} \\
& &   English & MiniZinc & Masters& CD& \href{https://www.uu.se/en/study/course?query=1DL451}{Modelling for Combinatorial Optimisation} \\
PT & Tecnico Lisboa & Portuguese & & & CD & \href{https://fenix.tecnico.ulisboa.pt/disciplinas/ALC/2023-2024/1-semestre}{Algoritmos para Lógica Computacional } \\
BE & Louvain & English & & & Video& \href{http://minicp.org}{MiniCP: A lightweight Constraint Programming Solver} \\
IE & University College Cork& English & Choco-solver & & CD & \href{https://ucc-ie-public.courseleaf.com/modules/}{Constraint Programming and Optimisation} \\
 & & English & lp-solve & Masters& CD & \href{https://ucc-ie-public.courseleaf.com/modules/}{Optimisation} \\
& & English & MiniZinc & PhD students& CD & \href{https://ucc-ie-public.courseleaf.com/modules/}{Artificial Intelligence Methods} \\
SE & Lund University& English & & & CD & \href{https://kurser.lth.se/lot/course-syllabus-en/23_24/EDAN01}{Constraint Programming} \\
FR & Lifeware, INRIA& French, English & SWI Prolog& & CD & \href{https://lifeware.inria.fr/wiki/Main/ConstraintModeling}{Modèles et Algorithmes à base de Contraintes pour l'Aide à la Décision} \\
 & & English & SWI Prolog& & CD& \href{https://lifeware.inria.fr/wiki/Main/LogicProgramming}{Relational Programming} \\
FR & INRA & French & & & Slides& \href{https://forgemia.inra.fr/degivry/enac}{ENAC course on graphs and networks} \\
NO & Simula &  & & & & \url{https://www.simula.no/education/courses/crash-course-machine-learning} \\
IT & Univ Bologna & English & MiniZinc & Masters& CD &  \href{https://www.unibo.it/en/teaching/course-unit-catalogue/course-unit/2023/479038}{Decision Making with Constraint Programming} \\
 & & English & MiniZinc & Masters & CD & \href{https://www.unibo.it/en/teaching/course-unit-catalogue/course-unit/2023/446597}{Combinatorial Decision Making and Optimization} \\
UK & York& English & & & CD & \href{https://www.york.ac.uk/students/studying/manage/programmes/module-catalogue/module/COM00159M/2022-23}{Constraint Programming} \\
FR & INRAE & French & Choco-solver & Masters& CD & \href{https://miat.inrae.fr/schiex/teaching.shtml}{Modèles et algorithmes pour les problèmes d'optimisation complexes de l'IA} \\
 & &  & Python  & & WB& \href{https://colab.research.google.com/drive/1RepKLvwqcA1kUVkwtPT2JBzdUwwQ2Ys0?usp=sharing}{VisualSudoku-e.ipynb} \\
 & &  & Python& & WB& \href{https://colab.research.google.com/drive/1ew7IceldcAhyZZ0bHvaHynZyM-s-ne0l?usp=sharing}{VisualSudoku.ipynb} \\
 & &  & Python& & WB& \href{https://colab.research.google.com/drive/1pSDTU-gC1avDahtqDIb96VEn-VDspPBW?usp=sharing}{Master DC1e.pynb} \\
 & &  & & & & \url{https://colab.research.google.com/drive/1tzw36ImzdtLWDrbmleEYRewDhRwmIixe?usp=sharing} \\
 & &  & Python & & WB& \href{https://colab.research.google.com/drive/1CheyRvt54za5trDMBf7A8yhgiuk9r4cb?usp=sharing}{VariableElimination.ipynb} \\
FR &Université Polytechnique Hauts-de-France &  & & & & \url{https://filesender.renater.fr/?s=download&token=c538ffa0-96fb-4156-843e-959eb8ba7af0} \\
DE & BTU Cottbus-Senftenberg& German & Choco-solver, ECLiPSe& Masters& CD & \href{https://www.b-tu.de/modul/12472}{Einführung in die Constraint-Programmierung} \\
TR & DOKUZ EYLUL UNIVERSITY, Izmir&  & OPL& & CD & \href{https://debis.deu.edu.tr/ders-katalog/2022-2023/eng/en_132_1513_1625.html}{Constraint Programming} \\
FR & Université Côte d’Azur& French & & & CD, Slides& \href{https://upinfo.univ-cotedazur.fr/s1/resolution-problemes/}{Résolution de problèmes} \\
US & Lincoln, Nebraska & English & & & & \href{https://cse.unl.edu/~choueiry/Constraint-Courses.html}{ Foundations of Constraint Processing} \\
LU & University of Luxembourg& English & MiniZinc& & Slides& \href{http://hyc.io/teaching/is1.html}{Intelligent System, Constraint Programming} \\
CA & University of Toronto& English & & Grad& CD & \href{http://www.cs.toronto.edu/~fbacchus/csc2512/}{Advanced Propositional Reasoning} \\
FR & Montpellier&  French & & & CD & \href{https://docs.google.com/spreadsheets/d/1kZt5TLQ7VFotJbuEDRAXSFoG-Y3U-rFAB8bYenpMP9Y/edit#gid=74570599}{HAI9010I Contraintes} \\
AU & Monash University & English & MiniZinc & Masters & CD & \href{https://handbook.monash.edu/2023/units/FIT5216}{Modelling discrete optimisation problems} \\
% & &  & & & & \href{}{} \\
\bottomrule
\end{longtable}
\end{landscape}

\clearpage
\section{Course Materials}
\label{sec:coursematerials}

Some responses to the survey gave links to detailed sets of resources for number of courses on CP and related topics, we give an overview in Table~\ref{tab:coursematerials}, and discuss each such course in turn below. The materials here are not an exhaustive list, but rather a list of courses for which further materials were available to the authors.

\begin{table}[htbp]
\caption{\label{tab:coursematerials}Course Materials}
{\scriptsize
\begin{sideways}
\begin{tabularx}{\textheight-1.2cm}{>{\hsize=0.7\hsize\linewidth=\hsize\RaggedRight}X>{\hsize=1.3\hsize\linewidth=\hsize\RaggedRight}Xllll}\toprule
Authors & Title & Year & Language & CP System & Length \\ \midrule
H. Simonis & \href{https://eclipseclp.org/ELearning/index.html}{ECLiPSe ELearning Course}\cite{Simonis2015} & 2009 & English & ECLiPSe\cite{DBLP:journals/tplp/SchimpfS12} & 20 chapters\\
Marie Pelleau & \href{https://upinfo.univ-cotedazur.fr/s1/resolution-problemes/}{Résolution de problèmes} &  & French & - & 5 chapters  \\
Pierre Talbot & \href{http://hyc.io/teaching/is1.html}{Intelligent System, Constraint Programming} & & English & - & 61 slides \\
Alexandre Gondran, Simon de Givry & \href{https://forgemia.inra.fr/degivry/enac}{ENAC course on graphs and networks} & 2021 & French, English & - & 273 slides \\
Claude-Guy Quimper & \href{http://www2.ift.ulaval.ca/~quimper/Optimisation/Diapositives/chapitres.php}{Optimization} & 2022 & French & - & 9 chapters\\
Pierre Flener & \href{https://user.it.uu.se/~pierref/courses/COCP/slides/}{Modelling for Combinatorial Optimisation} & 2023 & English & MiniZinc & 11 chapters\\
\bottomrule
\end{tabularx}
\end{sideways}
}
\end{table}

\subsection{ECLiPSe Elearning Course}
\label{sec:eclipseelearningcourse}

The "ECLiPSe Elearning Course" by H. Simonis consists of a series of lectures on using the ECLiPSe system for CP. The slides are available under an CC-NC-SA licence, and cover the following 20 lectures.

\begin{enumerate}
    \item Introduction (35 slides) \label{item:intro}
    \item First Steps 
    \item Constraint Application Overview (34 slides) \label{item:applications}
    \item Basic Constraint Reasoning - SEND+MORE=MONEY (70 slides) \label{item:basic}
    \item Global Constraints - Sudoku (87 slides) \label{item:global}
    \item Search Strategies - N-Queens (54 slides) \label{item:search}
    \item Optimization - Routing and Wavelength Assignment (RWA) (46 slides)
    \item Symmetry Breaking - BIBD (52 slides) \label{item:symmetry}
    \item Choosing the Model - Sports Scheduling (95 slides) \label{item:model}
    \item Customising Search - Progressive Party (64 slides) \label{item:customsearch}
    \item Limits of Propagation - Costas Array (50 slides)
    \item Systematic Development (28 slides)
    \item Visualization (54 slides) \label{item:visualization}
    \item Finite Set and Continuous Variables - SONET Design Problem (24 slides)
    \item Network Applications (85 slides)
    \item More Global Constraints - Car Sequencing (60 slides) \label{item:moreglobal}
    \item Using Mixed Integer Programming - RWA II (35 slides)
    \item A Hybrid Model - RWA III (47 slides) \label{item:hybrid}
    \item Comparing Technologies - RWA IV(52 slides)
    \item Working with Implications - Shikaku (57 slides)
\end{enumerate}

Most lectures introduce some concept based on a specific puzzle or problem, and provide a visualization of the problem and the solution process, based on the visualization tools described in~\cite{DBLP:conf/cp/SimonisDFMQC10}. Notable gaps in the course are scheduling and placement problems, these were not included as ECLiPSe did not possess the required global constraints. A Chapter on packing based on~\cite{Simonis2008, Simonis2011} using SICStus Prolog~\cite{carlsson2010sicstus} was used in some of the derived courses below, a Chapter on Methodology was added for the ACP Summer School in 2018, and the Chapters on visualization and applications have been replaced in the CRT-AI CP week with more up-to-date versions, as described in Section~\ref{sec:tutorials}. 

Selected elements of the course have been used for a number of courses, Summer Schools and presentations, typically only including some of the chapters of the overall course. Examples are:
\begin{description}
    \item[Doctoral School on Decision Theoretic AI, Cork, 2009] Chapters \ref{item:intro}, \ref{item:basic}, \ref{item:global}, \ref{item:search}, \ref{item:model}, \ref{item:applications}
    \item[ACP Summer School 2009, Cork] Chapters \ref{item:intro}, \ref{item:basic}, \ref{item:global}, \ref{item:search}, \ref{item:model}, \ref{item:applications}
    \item[CPAIOR 2009 Master Class] Chapters \ref{item:global}, \ref{item:model}, \ref{item:symmetry}, \ref{item:customsearch}, \ref{item:moreglobal}, \ref{item:hybrid}
    \item[JFPC 2010 Invited Talk] Chapters \ref{item:visualization}, \ref{item:global}, \ref{item:model}
    \item[Dagstuhl, Germany] Chapters \ref{item:intro}, \ref{item:basic}, \ref{item:global}, \ref{item:search}, \ref{item:customsearch}
    \item[CP and Verification, Izmir, Turkey]
    \item[ICON Summer School, 2015, Sicily, Italy] Chapters \ref{item:intro}, \ref{item:basic}, \ref{item:global}, \ref{item:search}, \ref{item:customsearch}
    \item[ACP Summer School 2016, Cork] Chapters \ref{item:intro}, \ref{item:basic}, \ref{item:global}, \ref{item:search}, \ref{item:model}, \ref{item:applications}
    \item[ACP Summer School 2018, Jackson, Wy] Chapters \ref{item:basic}, \ref{item:global}, \ref{item:search}, \ref{item:model}, \ref{item:customsearch}, \ref{item:moreglobal}
    \item[CRT-AI CP Week 2019, 2021, 2022, 2023, 2024] Chapters \ref{item:basic}, \ref{item:global}, \ref{item:search}, adapted to MiniZinc
\end{description}

\subsection{Résolution de problèmes}

The course "Résolution de problèmes" by Marie Pelleau of the Université Côté d’Azur in France consists of five
%\footnote{We need to check if these are all the slides of the course, or if other material is available.} 
lectures that deal with different aspects of combinatorial problem solving:
\begin{enumerate}
    \item Introduction (29 slides)
    \item Modelling (9 slides)
    \item Greedy Methods (21 slides)
    \item Local Search (25 slides)
    \item Constraint Programming (41 slides)
\end{enumerate}

\subsection{Graphs and Networks}
\label{sec:constraintsandnetworks}

The course "Graphs and Networks" by Alexandre Gondran and Simon de Givry for ENAC (Ecole Nationale de l'Aviation Civile) presents a lot of the material that one would need to know to understand the filtering methods used by important global constraints, while talking about constraint reasoning at the end. Main topics covered are:
\begin{enumerate}
\item Graphs and their properties
\item DFS, BFS, shortest path, MST
\item TSP, VRP
\item Matching
\item Graph Coloring
\item Network Flow
\item NP completeness
\item CSP
\end{enumerate}

\subsection{Combinatorial Optimization}

The "Combinatorial Optimization" course by Claude-Guy Quimper of the Université Laval in Quebec, Canada gives an overview of CP and Combinatorial Optimization, based on the following nine chapters:
\begin{enumerate}
    \item Introduction (54 slides)
    \item Search and Filtering (87 slides)
    \item Solver Design (44 slides)
    \item Filtering for Arithmetic Constraints (55 slides)
    \item Tractable Sub-classes (59 slides)
    \item Search Strategy (29 slides)
    \item Max Flow (87 slides)
    \item Linear Programming (131 slides)
    \item Conclusion (12 slides)
\end{enumerate}

This course is more focused on the problem solving aspect of CP, and does not discuss modelling or provide application examples.

\subsection{Modelling for Combinatorial Optimization}

The course "Modelling for Combinatorial Optimisation" by Pierre Flener at Uppsala University in Sweden consists of 8 modules, and is, in itself, part of a longer course on "Combinatorial Optimisation and Constraint Programming". The topics covered are
\begin{enumerate}
    \item Introduction (102 slides)
    \item Basic Modelling (61 slides)
    \item Constraint Predicates(53 slides)
    \item Modelling (22 slides)
    \item Symmetry (45 slides)
    \item Case Studies (34 slides)
    \item Solving Technologies (74 slides)
    \item Inference and Search (46 slides)
    % the following chapters have been removed as requeted by Pierre Flener
%    \item Modelling for Constraint Based Local Search (27 slides)
%    \item Modelling for SAT and SMT (22 slides)
%    \item Modelling for MIP (21 slides)
\end{enumerate}

The course uses MiniZinc and different back-end solvers to present the different concepts.

The longer course "Combinatorial Optimisation and Constraint Programming" extends the modelling course with on-site teaching of the material covered in the \href{https://www.edx.org/learn/computer-programming/universite-catholique-de-louvain-constraint-programming}{MiniCP MOOC}\cite{Delecluse2023, delecluse2023review} listed in Section~\ref{sec:minicp}.

\subsection{Constraint Programming: Star Wars Edition}

The course "Constraint Programming: Star Wars Edition" by Pascal Van Hentenryck and Tejas Santanam at Georgia Tech is a Star Wars-themed one-semester online course for both undergraduate and graduate students that is a modelling-focused course comprised of lecture videos, synchronous review sessions, and modeling project assignments. The course is a longer form course where each subtopic merits dozens of slides or hours of instruction and review. The topics covered in the course are listed below, and further information on the course can be found in the associated submission to The Workshop on Teaching Constraint Programming 2023~\cite{Santanam2023, santanam2023modern}.

\begin{enumerate}
    \item Basic Concepts
    \begin{itemize}
        \item Getting started
\item Basic concepts I
\item Basic concepts II
\item OPL Primer
    \end{itemize}
\item  Elements of Constraint Programming
\begin{itemize}
    \item Reified constraints
\item Optimization
\item Expressions
\end{itemize}
\item  Theoretical Foundation
\begin{itemize}
    \item Computational Model
    \end{itemize}
\item  Global Constraints
\begin{itemize}
    \item The element constraint
\item The table constraint
\item Combinatorial Constraints
\item The pack constraint
\item The circuit constraint
\item The lex constraints
\end{itemize}
\item  Modeling in Constraint Programming
\begin{itemize}
\item Symmetry breaking
\item Subexpression elimination
\item Redundant constraints I
\item Redundant constraints II
\end{itemize}
\item Search in Constraint Programming
\begin{itemize}
    \item Search tree and Impact
\item Restart and nogoods
\end{itemize}
\item Implementation of Constraint Programming
\begin{itemize}
\item Packing
\item AllDifferrent
\item NoOverlap
\end{itemize}
\item Scheduling in Constraint Programming
\begin{itemize}
\item Interval variables and noOverlap
\item The Sequence Constraints
\item Cumulative Constraints
\item The House Problem II
\item The House Problem III
\item The Perfect Square Problem
\item State Constraints
\item The Trolley Application
\item Optional Activities
\item  Standard Scheduling Problems
\item  Calendars
\end{itemize}
\item Advanced Topics
\begin{itemize}
\item  Large neighborhood search
\item  Scripting models
\item  Routing
\item  CP in Python
\end{itemize}
\item Implementation in MiniCP
\begin{itemize}
\item  Semantics of CP
\item  Operational Model of CP
\item  Inference
\item  Search
\item  Advanced Inference
\item  Advanced Search
\end{itemize}
\end{enumerate}

\clearpage
\section{Books}
\label{sec:books}

Table~\ref{tab:books} gives an overview of books that were or are currently by used in teaching Constraint Programming\footnote{Items in gray: I currently do not have a hard copy of this book to validate some entries}. Some of the entries are not intended as textbooks, they are either, like \cite{DBLP:books/daglib/0066904}, based on a PhD thesis, or can be seen, like \cite{DBLP:books/daglib/0017646}, as research monographs. But their content makes them suitable as a source of examples, or descriptions of constraints or algorithms, for courses on CP.
The "Handbook of Constraint Programming"~\cite{DBLP:reference/fai/2} was also mentioned as a reference by some of the survey participants, it consists of a series of chapters describing the state of research at time of publication, 2007.

The table gives the authors and title of the book, the year the book was published, and the number of pages. Two of the books~(\cite{DBLP:books/daglib/0024693} and \cite{DBLP:books/aw/RN2020}) are general introductions to Artificial Intelligence, we also list. the length of the chapters on Constraint Satisfaction. While the majority of books are written in English, there are a number of French books, and one book written in German. The English version of~\cite{Niederlinski2014} is a translation from a Polish original. We are not aware of books written, or translated, in other languages. We also list the CP system used in each of the books, some entries do not use an actual system in their description, while only one~\cite{Bourreau2019} has examples using multiple systems. We tried to classify the intended audience of the book, using information given in the books themselves, the classification used is
\begin{description}
    \item[U] undergraduate students
    \item[UU] advanced undergraduates
    \item[G] graduate students
    \item[R] researchers
    \item[M] masters programs
\end{description}

Finally, some of the books contain exercises, some even with solutions, which are particularly helpful for self-study.

\begin{table}[htbp]
\centering\ra{1.3}
\caption{Books for Learning Constraint Programming}
\label{tab:books}
{\scriptsize
\begin{sideways}
\begin{tabularx}{\textheight-1.2cm}{>{\hsize=1.0\hsize\linewidth=\hsize\RaggedRight}X >{\hsize=2.0\hsize\linewidth=\hsize\RaggedRight}X r r c c >{\hsize=0.4\hsize\linewidth=\hsize\RaggedRight}X >{\hsize=0.6\hsize\linewidth=\hsize\RaggedRight}X  c}
\toprule
Author & Title &  Year & Pages & Language & CP System & Audience & Topics &  Exercises \\ [0.5ex]
\midrule
P Van Hentenryck & Constraint satisfaction in logic programming\cite{DBLP:books/daglib/0066904}&1989 & 224 & English & CHIP\cite{DBLP:conf/fgcs/DincbasHSAGB88} & - & FD &  no\\
F. Fages & Programmation logique par contraintes\cite{Fages1998} & 1996 & 192 & French & GNU Prolog& U & Prolog, FD&  yes\\
K. Marriott, P. Stuckey & Programming with Constraints\cite{Marriott1998} &  1998 & 467 & English & CLP(R)\cite{DBLP:conf/compcon/JaffarMSY91} & UU & CLP(R)&  yes\\
P. Van Hentenryck & The {OPL} Optimization Programming Language\cite{opl1999} & 1999 & 254 & English & OPL\cite{DBLP:journals/informs/Hentenryck02} & - & FD&  no\\
J. Hooker & Logic-Based Methods for Optimization\cite{Hooker2000} & 2000 & 495 & English & - &  R, G & FD, MIP &  no \\
K. Apt & Principles of Constraint Programming\cite{DBLP:books/daglib/0018273} & 2003 & 407 & English & - & UU & Prolog, FD&  yes\\
\na R. Dechter & Constraint processing\cite{DBLP:books/daglib/0016622} & 2003 & 481 & English & - & ??? & CSP&  ???\\
T. Fr{\"{u}}hwirth, S. Abdennadher & Essentials of constraint programming\cite{DBLP:books/daglib/0008152} & 2003 & 156 & English & CHR & G,R & FD, CHR&  no\\
K. Apt, M. Wallace & Constraint Logic Programming using ECLiPSe \cite{DBLP:books/daglib/0018272}&  2007 & 329 & English & ECLiPSe\cite{DBLP:journals/tplp/SchimpfS12} & UU,G & Prolog, FD&  yes\\
J. Hooker & Integrated Methods for Optimization\cite{DBLP:books/daglib/0017646} & 2007 & 486 & English & - & R, G & FD, MIP&  yes\\
P. Hofstedt, A. Wolf & Einf{\"{u}}hrung in die Constraint-Programmierung\cite{DBLP:books/daglib/0034544} & 2007 & 388 & German & \Shortunderstack[l]{TURTLE\cite{Turtle} firstcs\cite{Wolf2012}}& M & FD &  yes\\
D. Poole, A. Mackworth & Artificial Intelligence - Foundations of Computational Agents\cite{DBLP:books/daglib/0024693} & 2010 & \Shortunderstack[r]{900 {(CSP ???)}}& English & - & UU, G& CSP & yes\\
\na C. Lecoutre & Constraint Networks: Targeting Simplicity for Techniques and Algorithms\cite{Lecoutre2013}&2013& 320& English & ??? & ??? & CSP, FD&  ???\\
A. Niederlinski& A Gentle Guide to Constraint Logic Programming via ECLiPSe\cite{Niederlinski2014} & 2014&509 & English & ECLiPSe\cite{DBLP:journals/tplp/SchimpfS12} & U & Prolog, FD&  yes \\
\na E. Tsang & Foundations of Constraint Satisfaction: The Classic Text\cite{Tsang2014} & 2014 & 444 & English & ??? & ??? & CSP&  ??? \\
N. Zhou, H. Kjellerstrand, J. Fruhman& Constraint Solving and Planning with Picat\cite{DBLP:series/sbis/ZhouKF15} & 2015 &140 & English & Picat\cite{DBLP:conf/ruleml/Zhou16} & U& FD&  yes\\
E. Bourreau, M. Gondran, P. Lacomme, M. Vinot& De la programmation linéaire à la programmation par contraintes\cite{Bourreau2019} & 2019 & 348& French & 
\Shortunderstack[l]{Gusek CPLEX GLPK Choco-solver\cite{DBLP:journals/jossw/PrudhommeF22}} & 
%\shortstack[l]{Gusek\\CPLEX\\GLPK\\Choco-solver\cite{DBLP:journals/jossw/PrudhommeF22}} & 
M& FD, MIP&  no\\
E. Bourreau, M. Gondran, P. Lacomme, M. Vinot& Programmation par Contraintes\cite{Bourreau2020} & 2020 & 232& French & Choco-solver\cite{DBLP:journals/jossw/PrudhommeF22} & M & FD&  no\\
S. Russell, P. Norvig & Artificial Intelligence: {A} Modern Approach (4th Edition)\cite{DBLP:books/aw/RN2020} & 2020 & \Shortunderstack[r]{1115 {(CSP 28)}}& English & - & U& CSP, FD&  no\\
M. Wallace & Building Decision Support Systems - using {MiniZinc}\cite{DBLP:books/sp/Wallace20} & 2020 & 224 & English & MiniZinc\cite{DBLP:conf/cp/NethercoteSBBDT07} & G & FD, MIP, SAT&  yes\\
\bottomrule
\end{tabularx}
\end{sideways}
}
\end{table}

\clearpage
\section{Video Courses}
\label{sec:videocourses}

Table~\ref{tab:videocourses} lists historical and current video courses on Constraint Programming. Where available, a link to the course itself is provided. We also show references describing the courses, the year the course was first made available, the language the course material is presented in, and the CP system that is used in the course. Finally, the length of the course is provided, either as the length of the video itself, or the length of the course expressed in hours, as stated by the course providers.

\begin{table}[htbp]
\caption{\label{tab:videocourses}Video Courses}
{\scriptsize
\begin{sideways}
\begin{tabularx}{\textheight-1.2cm}{>{\hsize=0.7\hsize\linewidth=\hsize\RaggedRight}X>{\hsize=1.3\hsize\linewidth=\hsize\RaggedRight}Xllll}\toprule
Authors & Title & Year & Language & CP System & Length \\ \midrule
M. Dincbas, H. Simonis & CHIP (Part I and II) \cite{Dincbas1989} & 1989 & English & CHIP\cite{DBLP:conf/fgcs/DincbasHSAGB88} & 2 hours\\
C. Solnon & Cours de Programmation Par Contraintes\cite{DBLP:journals/tsi/Solnon03,DBLP:conf/jfplc/Solnon03} & 2003 & French & \\
H. Simonis & \href{https://eclipseclp.org/ELearning/index.html}{ECLiPSe ELearning Course}\cite{Simonis2015} & 2009 & English & ECLiPSe\cite{DBLP:journals/tplp/SchimpfS12} & 22 Lectures\\
J. Lee, P. Stuckey & \href{https://www.coursera.org/learn/basic-modeling}{Basic Modeling for Discrete Optimization}\cite{DBLP:conf/aaai/ChanCFLS20,DBLP:conf/cp/Lee23,DBLP:conf/icbl/Lee21}& 2019 & English, Mandarin & MiniZinc\cite{DBLP:conf/cp/NethercoteSBBDT07} & 27 hours\\
& \href{https://www.coursera.org/learn/advanced-modeling}{Advanced Modeling for Discrete Optimization} & 2019 & English, Mandarin & MiniZinc\cite{DBLP:conf/cp/NethercoteSBBDT07} & 46 hours\\
& \href{https://www.coursera.org/learn/solving-algorithms-discrete-optimization}{Solving Algorithms for Discrete Optimization} & 2019 & English, Mandarin & MiniZinc\cite{DBLP:conf/cp/NethercoteSBBDT07} & 22 hours\\
\label{sec:minicp}P. Schaus, L. Michel, P.Van Hentenryck& \href{https://www.edx.org/learn/computer-programming/universite-catholique-de-louvain-constraint-programming}{LouvainX: Constraint Programming}\cite{Delecluse2023, delecluse2023review} & 2023 & English & MiniCP\cite{MiniCP2021} & \shortstack[l]{14 weeks\\6-8 hours per week}\\
\bottomrule

\end{tabularx}
\end{sideways}
}
\end{table}

\clearpage
\section{Articles}
\label{sec:articles}

A Workshop on Teaching Constraint Programming was run at CP 2015, but the proceedings of the workshop do no longer seem to be available on-line. Patrick Prosser organized the workshop, he also gave an invited talk at CP 2014\cite{DBLP:conf/cp/Prosser14}.

The \href{https://hsimonis.github.io/WTCP2023/}{Workshop on Teaching Constraint Programming 2023} at the CP 2023 conference in Toronto had six submissions~\cite{Lecoutre2023,Santanam2023,Choi2023,Cane2023,Alos2023,Delecluse2023, santanam2023modern, delecluse2023review}, an invited talk by Peter Stuckey and an overview presentation. The current paper is an extended version of that overview.

There has also been exploration into how people approach CP which would be useful to understand when trying to teach the subject. \cite{hoffmann_et_al:LIPIcs.CP.2022.28}

\clearpage
\section{Tutorials}
\label{sec:tutorials}

The tutorials covered in this section are shorter presentations, intended to give an overview of Constraint Programming, its applications or one of its major research topics. They can be presented as videos, and/or as slide sets. In Table~\ref{tab:tutorials} we show the authors, and the title of the tutorial, together with a link to an on-line version. We list the year the tutorial was made, the language it is presented in, and the CP system used. We differentiate between tutorials given as slide sets and videos. For the slides sets, we show the number of slides, for the videos the length of the video.

Every year, the Association for Constraint Programming (\href{https://www.a4cp.org/}{ACP}) organizes a Summer School on specific Constraint Programming topics. A list of the Summer Schools and the topics covered is presented at \href{https://www.a4cp.org/events/summer-schools}{ACP Summer Schools}. In some years, the lectures of the Summer School are made available outside the school, for example for the 2023 Summer School on Constraint Programming and Machine Learning, all lectures are available on YouTube \href{https://www.youtube.com/playlist?list=PLcByDTr7vRTYJ2s6DL-3bzjGwtQif33y3}{2023 ACP Summer School YouTube Channel}.

\begin{table}[htbp]
\caption{\label{tab:tutorials}Tutorials}
{\scriptsize
\begin{sideways}
\begin{tabularx}{\textheight-1.2cm}{lXllllr} \toprule
Authors & Title & Year & Language & CP System & Format & Length\\ \midrule
K. Kuchcinski & \href{https://www.lth.se/fileadmin/programvaruportal/2010/presentation2010-05-19.pdf}{Constraint Programming in Embedded System Design} & 2010 & English & MiniZinc~\cite{DBLP:conf/cp/NethercoteSBBDT07}, JaCoB~\cite{JaCoP2013} & Slides & 32 slides\\
H. Simonis & \href{https://hsimonis.github.io/WTCP2023/presentations/applications.pdf}{Optimization Applications} & 2023 & English & various & Slides & 132 slides \\
H. Simonis & \href{https://hsimonis.github.io/WTCP2023/presentations/logistics.pdf}{Logistics} & 2022 & English & various & Slides & 79 slides \\
C. Beck & \href{https://www.youtube.com/watch?v=yYxM7x657O4}{Supply Chain Optimization: An Operations Research Perspective} & 2021 & English & & Video & 56:23 min\\
H. Simonis, G. Tack & \href{https://youtu.be/AI-ZfQtMLAU?si=J8vP4aZ94WNCxD3H}{Visualization for Constraint Programming} & 2022 & English & MiniZinc~\cite{DBLP:conf/cp/NethercoteSBBDT07}, others& Video & 74:39 min\\
H. Simonis & \href{https://hsimonis.github.io/WTCP2023/presentations/methodology.pdf}{Constraint Programming Methodology} & 2018 & English & n/a & Slides & 36 Slides\\
\bottomrule
\end{tabularx}
\end{sideways}
}
\end{table}

\clearpage
\section{Other Material}
\label{sec:othermaterial}

% \emph{Other webpages which contain useful material, either overview pages, problem collections}

Gene Freuder maintains a \href{https://www.pearltrees.com/constraints/resources/id39817957}{collection of resources} about Constraint Programming. This includes links to courses, tutorial, books and other teaching material.

The constraint group at the Uppsala University in Sweden maintains \href{https://www.it.uu.se/research/group/optimisation/resources/constraint}{a list of resources} on constraint related topics.

The \href{https://sofdem.github.io/gccat/}{Global Constraint Catalog}~\cite{DBLP:journals/constraints/BeldiceanuCDP07} provides a very comprehensive description of global constraints, but also contains some exercises that can be used for teaching CP. These exercises fall into three categories, modelling, filtering techniques and specific global constraints (alldifferent, diffn, cumulative, cycle, element, and others).

\clearpage
\section{CP Course Formats}
\label{sec:cpcourseformats}

% \emph{We may skip this section for a first version of the paper, or put in straw-man/existing descriptions.}

In the survey, and the other sections of this paper, available material and what is currently missing has been identified. This next section considers course formats to fill different niches for various audiences, and which topics would be useful. The hope is that these formats can lead to sample curricula and shared material sets.

So far we have identified five possible course variants. 

\subsection{One-Hour Overview}
This describes a short one-hour overview of what CP is and what it is good for. The audience would be students of different program types, but possibly also the general public. This should be relatively easy to achieve, and could be the first shared material set to be produced. Suggested topics for a one-off short session would be:
\begin{itemize}
    \item Examples of real-world problems solved with CP
    \item Description of the main underpinnings of CP, including branch-and-prune and constraint propagation
    \item Comparison of CP versus MIP
    \item Example of how to construct a CP model for a small problem such as map coloring
\end{itemize}

\subsection{Short Course}
This course would be used inside a more general course (Logic Programming, Declarative programming, AI, Optimization, specific application areas), and would give an idea of the overall concepts used in CP, as well as showing some case studies highlighting success stories. We will need to identify how much time is available for the CP part, and what background knowledge can be assumed. 

\subsection{Modelling Course}
This would be a course that focuses on modelling skills, but skips deeper explanations of solving technology. It would be useful to still include a high-level explanation of how the CP system actually works, and why that results in limits of the problem sizes that can be handled.

\subsection{Tools and Techniques}
This would be a course that focuses on different technologies to solve CP problems (FD, SAT, SMT, MIP, MDD, CBLS), and which explains in principle how the reasoning inside the solver works. It would be nice if the course would also explain why some methods work best for some problems, and fail for others. 
This could be perhaps split into independent segments for each technology?

An possible alternative would be a course describing different academic and commercial tools for CP. This might in particular be of interest for practitioners, especially if the strengths and limitations of different tools for different application domains are clearly delineated.

\subsection{Algorithms and Constraints}

Many global constraints rely on specific algorithms to perform their reasoning, like bipartite matching or max-flow for alldifferent and cardinality constraints. This is largely independent of the actual constraint solver used, the algorithm is called at some point to deduce the possible propagation. An interesting course would present major global constraints with the mechanisms used in their reasoning, perhaps extended by an example of the use of the constraint is a real-world application. This is different from a solver design course, where the focus is on the overall representation of the constraints and variables, and their generic interaction.

The "Graphs and Networks" course described in Section~\ref{sec:constraintsandnetworks} could form the basis of such a course, using a selection of global constraints from~\cite{DBLP:journals/constraints/BeldiceanuCDP07}. The "ECLiPSe Elearning Course" described in Section~\ref{sec:eclipseelearningcourse} uses the format of linking a specific application with the introduction of the global constraints required to model the problem, and using visualization to show how the constraint propagation allows to solve the problem.  

\subsection{Solver Course}

This course explains how to design and build a solver for a specific constraint technology, most likely using finite domain reasoning. This course is intended for Computer Science students that need to know how a solver operates, and that should get some experience in writing code and running performance test experiments on a complex system.

This content might already be covered by the MiniCP~\cite{Delecluse2023, delecluse2023review} course.

\section{Conclusions}
\label{sec:conclusions}

In this paper we have presented the results of a survey on Teaching Constraint Programming, which we ran as part of the Workshop on Teaching Constraint Programming at CP 2023 in Toronto. We have explained the motivation and methodology used to the survey, and described the questions that we included in the questionnaire. We have presented the detailed answers for the questions, and performed a detailed analysis to understand what material is currently available and which topics are currently not well covered. We have also listed books, video courses, and course material collection that show how different aspects of CP are currently taught. Finally, we have briefly discussed which courses might be developed in the future, and which specific areas should be explored as a priority.

The goal of this paper and the hope is that discussions on teaching CP can be started and lead to the widespread open-source sharing of materials, course styles, and ideas to everyone from those teaching CP for the first time or to experienced CP educators. Any gaps in coverage within this paper is meant to illustrate a future opportunity rather than a blind spot.

This paper is also focused on courses in the academic sphere, though there is certainly merit in exploring CP use and education in industry. Unfortunately, the authors were not privy to complete information about CP education in industry. Despite that, CP is used today in industry for applications ranging from manufacturing to autonomous vehicles. A potential future area of exploration is the integration of CP education with industry to provide application-based projects that can serve to motivate students and increase job opporunities within CP in industry.

It also remains to be seen the impact that generative AI tools will have on CP education. As the CP community plans and designs curricula, exercises, and exams going forward, it is vital to strike the balance between a student learning the content and concepts with the help of generative AI tools like ChatGPT or a student mindlessly letting generative AI tools do their schoolwork for them.

Ultimately, the path toward continued growth for the CP community stems from the development of further CP practitioners and educators via CP courses. This paper attempts to serve as a snapshot and survey of the methods and resources in the field to make it easier to plan and teach CP or develop resources to that end.

\section*{Acknowledgment}
We want to thank all participants in the survey, and the authors and attendees of the WTCP23 workshop for their valuable view points and discussions. Special thanks to the Association for Constraint Programming and its president, David Bergman, for supporting this initiative.

\clearpage
\bibliographystyle{plain}
\bibliography{bib}

\appendix
\section{More Survey Results}
Given the rise of large language models in recent years, we asked about positive or negative experiences with such tools in the courses. The majority of replies answered "No", or "Not Yet", and some participants were confused by the question.
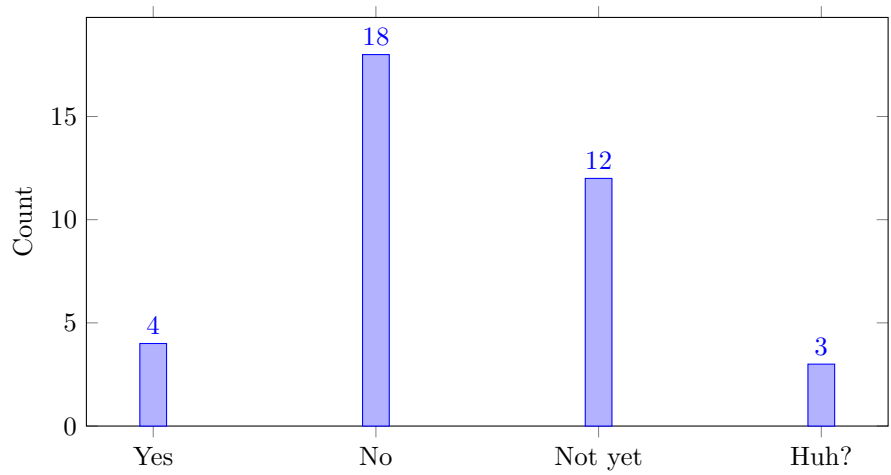
\begin{figure}[htbp]
\centering
\begin{tikzpicture}
\begin{axis}[ybar,symbolic x coords={Yes,No,Not yet,Huh?},width=\textwidth,height=7cm,ymin=0,xtick=data,ylabel=Count,nodes near coords, nodes near coords align={vertical}]
\addplot coordinates{(Yes,4)(No,18)(Not yet,12)(Huh?,3)};
\end{axis}
\end{tikzpicture}
\caption{\label{fig:llm}Have you seen any impact of Large Language Models (LLM)?}
\end{figure}
The "Yes" answers show that the impact does not have to be negative, but that the course will probably have to adapt to avoid cheating by students in their assignments.
\begin{itemize}
    \item Added material to course to show potential benefits
    \item Students use it to generate part of their coursework
    \item Some assignments can be solved by using ChatGPT
    \item A student found a bug in her program by asking ChatGPT for help
\end{itemize}

\end{document}